\documentclass[prb,superscriptaddress,reprint]{revtex4-1}

\usepackage{graphicx,bm,color}

\newcommand{\bfr}{{\bf r}}

\begin{document}

\title{The particle-hole map: formal derivation and numerical implementation}

\author{Yonghui Li}
\affiliation{Department of Physics, Tianjin University, 92 Weijin Road, Tianjin,
China, 300072}
\affiliation{Department of Physics and Astronomy, University of Missouri,
Columbia, Missouri, 65211}

\author{Carsten A. Ullrich}

\affiliation{Department of Physics and Astronomy, University of Missouri,
Columbia, Missouri, 65211}

\date{\today }

\begin{abstract}
The particle-hole map (PHM) is a tool to visualize electronic excitations, based on representations in a canonical orbital
transition space. Introduced as an alternative to the transition density matrix, the PHM has a simple probabilistic interpretation,
indicating the origins and destinations of electrons and holes and, hence,  the roles of different functional units of molecules
during an excitation.  We
present a formal derivation of the PHM, starting from the particle-hole transition density matrix and projecting onto
a set of single-particle orbitals. We implement the PHM using atom-centered localized basis sets and discuss the example of
the molecular charge-transfer complex C$_2$H$_4$--C$_2$F$_4$.
\end{abstract}

\maketitle

\section{Introduction}

Reduced density matrices (RDMs) have been extre\-me\-ly important and fruitful in the quantum mechanical description
of systems of interacting electrons. \cite{Cioslowski,Lowdin1955,McWeeny1960}  For instance, the one-particle
RDM gives rise to the natural orbitals. \cite{Davidson1972} Furthermore,
the ground-state energy of any $N$-electron system
can be expressed via the two-particle RDM, which opens up interesting possibilities for
calculating electron correlation. \cite{Mazziotti2012}

The one-particle transition density matrix (TDM) \cite{Furche2001,Etienne2015} plays an important role in the calculation and visualization of
electronic excitations using time-dependent density-functional theory (TDDFT).  \cite{Ullrich2012}
It has been frequently used to directly illustrate excitonic effects in conjugated polymers,
\cite{Mukamel1997,Tretiak2002,Tretiak2005}  because it provides a two-dimensional map of the excitation and allows the construction of
descriptors from which the exciton coherence and
delocalization lengths can be extracted.\cite{Plasser2012,Bappler2014,Plasser2014,Mewes2015} Furthermore, the one-particle TDM is the basis for constructing
the natural transition orbitals. \cite{Martin2003}

However, in spite of its appeal, the TDM has some disadvantages, which motivated us to search for an alternative.\cite{Li2011,Li2015}
First of all, the TDM is, in general, complex (being related to the exciton wave function\cite{Plasser2012,Bappler2014,Plasser2014,Mewes2015}); for the standard frequency-dependent TDM, this is not an issue because one can choose real orbitals or use other techniques to extract useful information,
but for the time-dependent version of the TDM, \cite{Li2011} this introduces some degree of ambiguity how it should be represented.

Secondly, the physical interpretation of the TDM is such that it gives access to electron and hole distributions
of an excitation. This is extremely useful to characterize excitons; however, it does not tell us where the electron and the hole came
from, which is vital for understanding and interpreting excitation processes.
In other words, the information provided by the TDM is incomplete.

Lastly, the TDM in its common representation as a two-dimensional map\cite{Mukamel1997,Tretiak2002,Tretiak2005} can be
a bit inconvenient to read, because the exciton coherence and delocalization lengths are defined with respect to the diagonal of the map, rather than
with respect to horizontal and vertical directions.

We recently introduced a new concept, the particle-hole map (PHM),\cite{Li2015,Li2016}
which offers new and useful ways of visualizing charge motions and coherences during excitation processes in molecules
and solids. In simple words, the PHM tells us precisely where electrons and holes are coming from and where they are going
to, which allows us to identify the role of different functional units. This provides chemical insight about charge-transfer processes
which would be difficult to obtain with other methods.

In Ref. \onlinecite{Li2015}, we directly compared the PHM with the TDM
for simple one-dimensional model systems and found that the PHM plots were generally easier to read and interpret.
However, we emphasize that the TDM remains important in its own right because it is very well suited to characterize
the coherence of excitonic wave functions in large (or periodic) systems.

The PHM is not a physical observable in the sense that there is a ``PHM operator''.
This is not necessarily a drawback, since there exist other very successful visualization methods that are
not physical observables either, such as the electron localization function.\cite{Becke1990,Burnus2005}
Nevertheless, a more solid formal underpinning of the PHM, bringing it as closely as possible to a physical observable, would be desirable.

The main purpose of this paper is to present a formal derivation of the PHM for the $n$th excitation,
$\Xi(\bfr,\bfr',\Omega_n)$, starting from a particle-hole reduced TDM,
and using projections onto a set of canonical orbitals. This gives a complementary perspective to our earlier
derivation, \cite{Li2015} which was motivated by physical and probabilistic arguments based on orbital densities and density fluctuations.
The new derivation has the additional advantage that it can be formally made exact, involving many-body wave functions
and associated Dyson orbitals.

We have implemented the PHM as a practical tool for standard quantum chemistry codes such as  {\tt Gaussian} \cite{g09}
(our earlier implementation \cite{Li2015,Li2016} had been for the grid-based {\tt octopus} code\cite{octopus}).
 Using the example of the
$\rm C_2H_4$--$\rm C_2F_4$ charge-transfer complex, we
demonstrate that the PHM provides useful information to reveal the inner mechanisms of electronic excitation processes.

\section{Derivation of the PHM} \label{sec:theory}

\subsection{Reduced density matrices: basic definitions}

The $n$th order RDM $\gamma_n$ can be defined as follows: \cite{Mazziotti2012}
\begin{equation} \label{gamma_n}
\gamma_n = \langle \Psi | \hat C^\dagger \hat C | \Psi\rangle,
\end{equation}
where $\Psi$ is a many-body wave function and $\hat C$ denotes
the product of $n$ creation and/or annihilation operators (here and in the following, spin is ignored).
The easiest case is $n=1$, where
\begin{eqnarray} \label{gamma_1_p}
\gamma_1^p(\bfr,\bfr') &=& \langle \Psi | \hat \psi^\dagger (\bfr') \hat\psi(\bfr) | \Psi\rangle\\
\gamma_1^h(\bfr,\bfr') &=& \langle \Psi | \hat \psi (\bfr') \hat\psi^\dagger(\bfr) | \Psi\rangle \label{gamma_1_h}
\end{eqnarray}
are the one-particle and one-hole RDMs in real-space representation;
$\hat\psi(\bfr)$ and $\hat \psi^\dagger(\bfr)$ are fermionic field operators which, respectively, destroy and create
a particle at position $\bfr$. In terms of
the standard creation and destruction operators $\hat a_j^\dagger$ and $\hat a_j$ and an orbital basis, we have
\begin{equation} \label{orbital_basis}
\hat \psi(\bfr) = \sum_j \varphi_j(\bfr) \hat a_j \:, \quad
\hat \psi^\dagger(\bfr) = \sum_j \varphi_j^*(\bfr) \hat a_j^\dagger \:.
\end{equation}
Let us consider the case of a noninteracting system in its nondegenerate ground state, where $\Psi=\Phi_0$, a single
ground-state Slater determinant. One then obtains
\begin{eqnarray} \label{gamma_10_p}
\gamma_1^p(\bfr,\bfr') &=& \sum_{i}^{occ} \varphi_i(\bfr) \varphi_i^*(\bfr')\\
\gamma_1^h(\bfr,\bfr') &=& \sum_{a}^{unocc} \varphi_a(\bfr') \varphi_a^*(\bfr) \:. \label{gamma_10_h}
\end{eqnarray}
In Eq. (\ref{gamma_10_p}), the summation runs over all occupied orbitals, and in Eq. (\ref{gamma_10_h}) it runs over
all unoccupied orbitals.

There are four possibilities for the second-order case: we can destroy or create two particles or a particle-hole pair.
We thus obtain the following RDMs:
\begin{eqnarray}
\gamma_2^{pp}(\bfr_1,\bfr_2;\bfr_1',\bfr_2') &=&
\langle \Psi | \hat\psi^\dagger(\bfr_1') \hat \psi^\dagger(\bfr_2') \hat \psi(\bfr_2) \hat\psi(\bfr_1) | \Psi\rangle \\
\gamma_2^{ph}(\bfr_1,\bfr_2;\bfr_1',\bfr_2') &=&
\langle \Psi | \hat\psi^\dagger(\bfr_1') \hat \psi(\bfr_2') \hat \psi^\dagger(\bfr_2) \hat\psi(\bfr_1) | \Psi\rangle \\
\gamma_2^{hp}(\bfr_1,\bfr_2;\bfr_1',\bfr_2') &=&
\langle \Psi | \hat\psi(\bfr_1') \hat \psi^\dagger(\bfr_2') \hat \psi(\bfr_2) \hat\psi^\dagger(\bfr_1) | \Psi\rangle \\
\gamma_2^{hh}(\bfr_1,\bfr_2;\bfr_1',\bfr_2') &=&
\langle \Psi | \hat\psi(\bfr_1') \hat \psi(\bfr_2') \hat \psi^\dagger(\bfr_2) \hat\psi^\dagger(\bfr_1) | \Psi\rangle \hspace{3mm}
\end{eqnarray}
For the noninteracting case this becomes
\begin{eqnarray}
\gamma_2^{pp}(\bfr_1,\bfr_2;\bfr_1',\bfr_2') &=&
\gamma_1^p(\bfr_1,\bfr_1')\gamma_1^p(\bfr_2,\bfr_2') \nonumber\\
&-&
\gamma_1^p(\bfr_2,\bfr_1')\gamma_1^p(\bfr_1,\bfr_2') \\
\gamma_2^{ph}(\bfr_1,\bfr_2;\bfr_1',\bfr_2') &=&
\gamma_1^p(\bfr_1,\bfr_1')\gamma_1^h(\bfr_2',\bfr_2) \nonumber\\
&+&
\gamma_1^p(\bfr_1,\bfr_2)\gamma_1^p(\bfr_2',\bfr_1') \\
\gamma_2^{hp}(\bfr_1,\bfr_2;\bfr_1',\bfr_2') &=&
\gamma_1^h(\bfr_1',\bfr_1)\gamma_1^p(\bfr_2,\bfr_2') \nonumber\\
&+&
\gamma_1^h(\bfr_2,\bfr_1)\gamma_1^h(\bfr_1',\bfr_2') \\
\gamma_2^{hh}(\bfr_1,\bfr_2;\bfr_1',\bfr_2') &=&
\gamma_1^h(\bfr_1',\bfr_1)\gamma_1^h(\bfr_2',\bfr_2) \nonumber\\
&-&
\gamma_1^h(\bfr_1',\bfr_2)\gamma_1^h(\bfr_2',\bfr_1) \:.
\end{eqnarray}

\subsection{First-order transition density matrices}

We define $n$th order TDMs by generalizing Eq. (\ref{gamma_n}):\cite{McWeeny1960}
\begin{equation} \label{tgamma_n}
\gamma_n(p,q) = \langle \Psi_p| \hat C^\dagger \hat C | \Psi_q\rangle.
\end{equation}
Here, $\Psi_p$ and $\Psi_q$ are two different $N$-particle wave functions. In the following,
one of the wave functions is the ground state and the other one is an excited state.

Let us again consider the noninteracting case. The first-order TDMs between $\Phi_0$
and an excited state $\Phi_{ia}$ (associated with a single-particle transition $i\to a$) are
\begin{eqnarray} \label{tgamma_1_p1}
\gamma_1^p(ia,0;\bfr,\bfr') &=&
\langle \Phi_{ia} | \hat \psi^\dagger(\bfr') \hat \psi(\bfr) | \Phi_0\rangle
=  \varphi_i(\bfr)\varphi_a^*(\bfr') \\
\gamma_1^p(0,ia;\bfr,\bfr') &=&
\langle \Phi_0 | \hat \psi^\dagger(\bfr') \hat \psi(\bfr) | \Phi_{ia}\rangle
= \varphi_i^*(\bfr') \varphi_a(\bfr)  \hspace{5mm}
\label{tgamma_1_p2}
\end{eqnarray}
and similar for $\gamma_1^h$.
All first-order TDMs between the ground state and multiply excited configurations vanish.

In TDDFT, the Kohn-Sham TDM associated with the $n$th excitation is obtained as a superposition of the TDMs associated with
individual single-particle transitions:\cite{Furche2001,Tretiak2002}
\begin{eqnarray} \label{Gamman}
\Gamma_n(\bfr,\bfr') &=&
 \sum_{ia} [ \gamma_1^p(ia,0;\bfr,\bfr') X_{ia}^{(n)}
+\gamma_1^p(0,ia;\bfr,\bfr') Y_{ia}^{(n)}] \nonumber\\
&=&
\sum_{ia} [ \varphi_i(\bfr) \varphi_a^*(\bfr') X_{ia}^{(n)}
+ \varphi_i^*(\bfr') \varphi_a(\bfr) Y_{ia}^{(n)}]
\end{eqnarray}
where $X_{ia}^{(n)},Y_{ia}^{(n)}$ are the eigenvectors of the Casida equation.\cite{Casida1995,Ullrich2012}
This result can also be obtained from a time-dependent perspective. Let $\Phi_t$ be a time-dependent
Slater determinant which evolves from $\Phi_0$.
We have
\begin{equation} \label{gamma_1t}
\gamma_1^p(\bfr,\bfr',t) = \langle \Phi_t | \hat\psi^\dagger(\bfr')\hat\psi(\bfr)|\Phi_t\rangle \:.
\end{equation}
Let us assume that the time evolution is in response to a weak perturbation:
$\Phi_t = \Phi_0 e^{-iE_0 t} + \delta \Phi_t$,
where $E_0$ is the ground-state energy. We expand
\begin{equation}
\delta \Phi_t = \sum_j C_j(t) e^{-iE_j t} \Phi_j \:,
\end{equation}
where the sum runs over all excited configurations. Substituting this into Eq. (\ref{gamma_1t}) we get, to first order,
\begin{eqnarray} \label{dgp1t}
\lefteqn{\delta \gamma_1^p(\bfr,\bfr',t) = } \\
&&
\sum_{ia}\Big[ C_{ia}^*(t)e^{i \omega_{ai} t} \varphi_i(\bfr)\varphi_a^*(\bfr')
+
C_{ia}(t)e^{-i \omega_{ai} t} \varphi_i^*(\bfr')\varphi_a(\bfr)\Big] \nonumber
\end{eqnarray}
where $\omega_{ai} = E_{ia} - E_0 = \varepsilon_a - \varepsilon_i$.
Notice that for $\bfr = \bfr'$ this reduces to the density response:
$\delta \gamma_1^p(\bfr,\bfr,t)=
\delta n(\bfr,t)$.
Now assume that the system is in an eigenmode corresponding to the $n$th excitation, with frequency $\Omega_n$.
The density response is then $\delta n(\bfr,t) =  e^{-i\Omega_n t} \delta n(\bfr,\Omega_n)$, and
\begin{equation}
\delta n(\bfr,\Omega_n) =
\sum_{ia} [ \varphi_i(\bfr) \varphi_a^*(\bfr) X_{ia}^{(n)}
+ \varphi_i^*(\bfr) \varphi_a(\bfr) Y_{ia}^{(n)}]
\end{equation}
Since $\delta \gamma_1^p(\bfr,\bfr',t) = e^{-i\Omega_n t}\delta \gamma_1^p(\bfr,\bfr',\Omega_n)$ in the single-mode case,
comparison of expressions (\ref{Gamman}) and (\ref{dgp1t})
allows us to identify the Fourier transforms of the coefficients $C_{ia}^*(t) e^{i \omega_{ai}t}$ and $C_{ia}(t) e^{-i \omega_{ai}t}$
with the eigenvectors $X_{ia}^{(n)}$ and $Y_{ia}^{(n)}$ of the Casida equation.

\subsection{Second-order transition density matrices and construction of the PHM}

Let us now consider the noninteracting second-order TDMs. We obtain, for single excitations,
\begin{eqnarray}
\lefteqn{
\gamma_2^{pp}(ia,0;\bfr_1,\bfr_2;\bfr_1',\bfr_2') } \label{gamma2t_pp}\\
&=&
\varphi_i(\bfr_1)\varphi^*_a(\bfr_1') \gamma_1^p(\bfr_2,\bfr_2')   + \gamma_1^p(\bfr_1,\bfr_1')  \varphi_i(\bfr_2)\varphi^*_a(\bfr_2') \nonumber\\
&-&
\varphi_i(\bfr_2)\varphi^*_a(\bfr_1') \gamma_1^p(\bfr_1,\bfr_2')-\gamma_1^p(\bfr_2,\bfr_1') \varphi_i(\bfr_1)\varphi^*_a(\bfr_2') \nonumber\\
\lefteqn{
\gamma_2^{ph}(ia,0;\bfr_1,\bfr_2;\bfr_1',\bfr_2') } \label{gamma2t_ph}\\
&=&
\varphi_i(\bfr_1)\varphi^*_a(\bfr_1') \gamma_1^h(\bfr_2',\bfr_2)-\gamma_1^p(\bfr_1,\bfr_1') \varphi_i(\bfr_2') \varphi^*_a(\bfr_2) \nonumber\\
&+&
\varphi_i(\bfr_1) \varphi^*_a(\bfr_2)\gamma_1^p(\bfr_2',\bfr_1')+ \gamma_1^p(\bfr_1,\bfr_2)\varphi_i(\bfr_2')\varphi^*_a(\bfr_1')  \nonumber\\
\lefteqn{
\gamma_2^{hp}(ia,0;\bfr_1,\bfr_2;\bfr_1',\bfr_2') } \label{gamma2t_hp}\\
&=&
\gamma_1^h(\bfr_1',\bfr_1)\varphi_i(\bfr_2)\varphi^*_a(\bfr_2')-  \varphi_i(\bfr_1')\varphi^*_a(\bfr_1)\gamma_1^p(\bfr_2,\bfr_2') \nonumber\\
&-&
\varphi_i(\bfr_2) \varphi^*_a(\bfr_1)\gamma_1^h(\bfr_1',\bfr_2')- \gamma_1^h(\bfr_2,\bfr_1)\varphi_i(\bfr_1') \varphi^*_a(\bfr_2')\nonumber\\
\lefteqn{
\gamma_2^{hh}(ia,0;\bfr_1,\bfr_2;\bfr_1',\bfr_2') } \label{gamma2t_hh}\\
&=&
- \varphi_i(\bfr_1') \varphi^*_a(\bfr_1) \gamma_1^h(\bfr_2',\bfr_2)- \gamma_1^h(\bfr_1',\bfr_1) \varphi_i(\bfr_2') \varphi^*_a(\bfr_2)
 \nonumber\\
&+&
\gamma_1^h(\bfr_1',\bfr_2) \varphi_i(\bfr_2')  \varphi^*_a(\bfr_1) +\varphi_i(\bfr_1')\varphi^*_a(\bfr_2) \gamma_1^h(\bfr_2',\bfr_1).\nonumber
\end{eqnarray}
The adjoined TDMs, such as $\gamma_2^{pp}(0,ia;\bfr_1,\bfr_2;\bfr_1',\bfr_2') $, are obtained by interchanging primed and unprimed
coordinates and complex conjugation, see  Eqs. (\ref{tgamma_1_p1}) and (\ref{tgamma_1_p2}).
The second-order TDMs for  doubly excited configurations
involve products of two occupied and two unoccupied orbitals; they will not contribute in the following.

To derive the PHM of the Kohn-Sham system,
let us define the projector onto the $l$th single-particle orbital:
\begin{equation}
P_l(\bfr,\bfr') = \varphi_l(\bfr) \varphi_l^*(\bfr').
\end{equation}
Applying this projector to the particle-hole TDM (\ref{gamma2t_ph}),
and summing over all occupied orbitals, gives
\begin{eqnarray} \label{proj}
\lefteqn{\hspace{-1.5cm}
\sum_l^{occ}\int \!\! \int  P_l^*(\bfr_2',\bfr_3') \gamma_2^{ph}(ia,0;\bfr_1,\bfr_2;\bfr_1',\bfr_2')
P_l(\bfr_2,\bfr_3) dr_2 dr_2'} \nonumber\\
&\equiv&
\sum_l^{occ} \langle P_l |  \gamma_2^{ph}(ia,0) | P_l \rangle (\bfr_1,\bfr_1';\bfr_3',\bfr_3) \nonumber\\
&=&
\varphi_i(\bfr_1)\varphi^*_a(\bfr_1') \varphi_i(\bfr_3')\varphi_i^*(\bfr_3) \:.
\end{eqnarray}
The projected hole-particle TDM (\ref{gamma2t_hp}), after a permutation of the coordinates,
gives the same result, apart from a minus sign.
On the other hand, the particle-particle (after coordinate permutation)
and hole-hole TDMs, (\ref{gamma2t_pp}) and (\ref{gamma2t_hh}), give zero, as do all doubly excited TDMs.

Setting $\bfr_1 = \bfr_1' = \bfr'$ and $\bfr_3 = \bfr_3' = \bfr$ in Eq. (\ref{proj}) gives
\begin{equation} \label{proj_same}
\sum_l^{occ} \langle P_l |  \gamma_2^{ph}(ia,0) | P_l \rangle (\bfr';\bfr)
= \varphi_i(\bfr')\varphi^*_a(\bfr') |\varphi_i(\bfr)|^2 \:.
\end{equation}
Now let us follow the time-dependent derivation of the first-order TDM given above. We have
\begin{equation} \label{gamma_2t}
\gamma_2^{ph}(\bfr_1,\bfr_2;\bfr_1',\bfr_2',t) =
\langle \Phi_t| \hat\psi^\dagger(\bfr_1')\hat\psi(\bfr_2')\hat\psi^\dagger(\bfr_2)\hat\psi(\bfr_1)|\Phi_t\rangle \:.
\end{equation}
Assuming that the time evolution has the form of a weak perturbation relative to $\Phi_0$,
the first-order term in $\gamma_2^{ph}$ is
\begin{eqnarray}
\lefteqn{\hspace{-0.9cm} \delta \gamma_2^{ph}(\bfr_1,\bfr_2;\bfr_1',\bfr_2',t)} \nonumber\\
&=& \langle \delta\Phi_t |\hat\psi^\dagger(\bfr_1')\hat\psi(\bfr_2')\hat\psi^\dagger(\bfr_2)\hat\psi(\bfr_1) |\Phi_0 \rangle e^{-i E_0 t}
\nonumber\\
&& {}+
\langle \Phi_0 | \hat\psi^\dagger(\bfr_1')\hat\psi(\bfr_2')\hat\psi^\dagger(\bfr_2)\hat\psi(\bfr_1)|\delta\Phi_t\rangle  e^{i E_0 t} \nonumber\\
&=&
\sum_{ia}\Big[ C_{ia}^*(t)e^{i \omega_{ai} t} \gamma_2^{ph}(ia,0;\bfr_1,\bfr_2;\bfr_1',\bfr_2')
\nonumber\\
&& {}+
C_{ia}(t)e^{-i \omega_{ai} t} \gamma_2^{ph}(0,ia;\bfr_1,\bfr_2;\bfr_1',\bfr_2')\Big] \nonumber\\
&+&
\mbox{\em doubly excited configurations.}
\end{eqnarray}
Carrying out the projection, and using Eq. (\ref{proj_same}), gives
\begin{eqnarray}
\lefteqn{
\sum_l^{occ} \langle P_l | \delta \gamma_2^{ph}(t)| P_l\rangle (\bfr';\bfr)} \nonumber\\
&=&
\sum_{ia}\Big[ C_{ia}^*(t)e^{i \omega_{ai} t} \varphi_i(\bfr')\varphi^*_a(\bfr') |\varphi_i(\bfr)|^2
+ c.c.  \Big]. \label{40}
\end{eqnarray}
Let the system be in an eigenmode corresponding to the $n$th excitation, and take the Fourier transform.
We then obtain, using the same arguments as above,
\begin{eqnarray}
\Xi(\bfr,\bfr',\Omega_n)
&=&
\sum_l^{occ} \langle P_l | \delta \gamma_2^{ph}(\Omega_n)| P_l\rangle (\bfr';\bfr)
 \nonumber\\
&=&
\sum_{ia}[  \varphi_i(\bfr')\varphi^*_a(\bfr') |\varphi_i(\bfr)|^2 X_{ia}^{(n)}
\nonumber\\
&& {}+
\varphi_i^*(\bfr')\varphi_a(\bfr') |\varphi_i(\bfr)|^2  Y_{ia}^{(n)} ]. \label{PHM}
\end{eqnarray}
This is the frequency-dependent Kohn-Sham PHM, which we had previously proposed using
intuitive physical arguments.\cite{Li2015,Li2016}
Notice that the PHM satisfies two important sum rules:
$\int \Xi(\bfr,\bfr',\Omega_n) d^3r' = 0$ (charge conservation) and
$\int \Xi(\bfr,\bfr',\Omega_n) d^3r = \delta n(\bfr',\Omega_n)$ (integration to the transition density).

\subsection{Discussion}

The PHM  (\ref{PHM}) is written as  the sum of probability density
fluctuations relative to the $i$th orbital, taken at position $\bfr'$, weighted by
the probability density of the $i$th orbital at position $\bfr$. This allows us to analyze the charge dynamics
during an excitation process, namely, it tells us the locations $\bfr$ where electrons and holes are
coming from in the ground state, and the locations $\bfr'$ where they are going to during the excitation.
As we now see, this interpretation of the PHM
is justified because of its construction via the particle-hole TDM.

The PHM can be compared to other existing methods in the literature which provide a visualization
of electrons and holes during an excitation process. The first-order TDM, which
we discussed above in Section II.B, can be regarded as the electron-hole amplitude;
in the limit of extended systems this becomes the exciton wave function.\cite{Plasser2014}
From the square of the TDM one can obtain electron and hole densities;
in Kohn-Sham approximation, these are given by
\begin{eqnarray}
\rho_H^{(n)}(\bfr) &=& \int d^3 r' |\Gamma_n(\bfr,\bfr')|^2, \\
\rho_E^{(n)}(\bfr') &=& \int d^3 r |\Gamma_n(\bfr,\bfr')|^2.
\end{eqnarray}
It is straightforward to see that both are normalized to $\sum_{ia}(|X_{ia}^{(n)}|^2 + |Y_{ia}^{(n)}|^2)$, which
in general is not equal to unity. Other local quantities, similar to $\rho_H^{(n)}$ and $\rho_E^{(n)}$, are the so-called attachment and detachment densities, which follow from the difference density matrix.\cite{Head1995}

A nonlocal visualization of the distribution of electrons and holes can be achieved via the so-called
electron-hole correlation plots: \cite{Plasser2014,Sun2006,Scheblykin2007}
\begin{equation}
\Omega_{AB}^{(n)} =  \int_A d^3r \int_B d^3r'  |\Gamma_n(\bfr,\bfr')|^2 \:.
\end{equation}
Here, $A$ and $B$ refer to spatial regions containing a specific atom, or a fragment, of the system.
$\Omega_{AB}^{(n)}$ is interpreted as the probability to find the hole on fragment $A$ and at the same time
the electron on fragment $B$.\cite{Plasser2014}
Notice that this is quite different from the information conveyed by the PHM:
the PHM ties the location of an electron or a hole at a specific position $\bfr'$ to a given ground-state point of origin $\bfr$.
It therefore correlates the electron and hole only indirectly to each other.

We illustrated the differences between the PHM and the one-particle TDM in Ref. \onlinecite{Li2015} using simple 1D model
systems, where the excitations do not have excitonic character. In these examples, as well as in all other systems that
we studied,\cite{Li2015,Li2016} the PHM turned out to provide a very clear and useful picture of the excitation mechanisms.
In general, for some systems the details and aspects of
electronic excitation processes may be better visualized with the PHM, in other situations the TDM may be more suitable.
Hence,  the PHM should be a useful addition to the toolbox of computational chemistry.

\subsection{Definition of the exact PHM}

The derivation of the PHM
presented in this section was carried out in terms of Kohn-Sham Slater determinants. It is well established that the Kohn-Sham orbitals, if obtained from high-quality exchange-correlation functionals,
are physically very meaningful objects, in the sense that they provide
excellent single-orbital descriptions of bound electrons and electronic excitations. \cite{Baerends2013,Meer2014} Furthermore,
the derivation of the PHM can be formally ``exactified'', as we now explain.

The first step, the construction of the exact second-order particle-hole TDM $\gamma_{\rm 2,ex}^{ph}$, can
be carried out as in Eq. (\ref{gamma_2t}), but using the exact time-dependent $N$-particle wave function of the system, $\Psi_t$.
The second step involves projection of the exact linearized TDM $\delta\gamma_{\rm 2,ex}^{ph}$ onto a set of single-particle orbitals.
The obvious choice in the context of a wave function based approach is to project onto the  Dyson orbitals
\cite{Katriel1980,Gritsenko2003,Gritsenko2016} of the $N$-particle system.

Dyson orbitals are formally defined as the overlap amplitudes between the $N$-particle and the $(N-1)$-particle system,
$d_l(\bfr) = \langle \Psi_l^{N-1}|\Psi_0^N(\bfr)\rangle_{N-1}$. In general, they are linearly dependent, nonorthogonal, not
normalized to unity, and do not have occupation numbers that are simply 0 or 1. Kohn-Sham orbitals can be viewed as
the Dyson orbitals of the noninteracting Kohn-Sham system.

Dyson orbitals belonging to so-called primary ionizations (which only involve the removal of one orbital, and no additional excitation or shake-up)
are known to have a norm close to 1 and to be very similar to the exact occupied Kohn-Sham orbitals,
but there are also other Dyson orbitals (corresponding to ``satellites''); these, however,
tend to have a norm much smaller than 1.\cite{Gritsenko2003,Gritsenko2016}
Hence, the projection step analogous to Eq. (\ref{40}) will now involve an infinite sum, but will be dominated by the
$N$ primary ionization Dyson orbitals. We end up with our definition of the exact PHM:
\begin{equation}
\Xi_{\rm ex}(\bfr,\bfr',\Omega_n)
=
\sum_l^{\infty} \langle P^{\rm dy}_l | \delta \gamma_{\rm 2,ex}^{ph}(\Omega_n)| P^{\rm dy}_l\rangle (\bfr';\bfr)\:,
\end{equation}
where $P^{\rm dy}_l = d_l(\bfr) d_l^*(\bfr')$ is the projector onto the $l$th Dyson orbital.


\section{Implementation and example}

\subsection{Graphical representation of the PHM}

For a given excitation $\Omega_n$, the PHM $\Xi(\bfr,\bfr',\Omega_n)$ is defined as a function of six spatial coordinates, and
there are various ways in which it can be graphically represented. In Ref. \onlinecite{Li2015} we introduced a spatial
partitioning scheme, to be used for real-space grid based codes such as {\tt octopus}. \cite{octopus} However,
the majority of electronic structure codes in quantum chemistry work with atom-centered basis sets.
Let us write the Kohn-Sham orbitals as
\begin{equation} \label{eq3.16}
\varphi_j(\bfr) = \sum_{l=1}^{N_A} \sum_{k=1}^{M_l} \zeta_{jk}^l \eta_k^l(\bfr) \:,
\end{equation}
where $N_A$ is the number of atoms in the molecule and $M_l$ is the number of basis functions in the $l$th atom.
If this expansion is inserted into Eq. (\ref{PHM}), one gets
cross-terms between basis functions belonging to different atoms, which introduces some complications. It is convenient
to make the approximation of zero overlap
in constructing the PHM, $\eta_k^l(\bfr)\eta_{k'}^{l'}(\bfr)\approx 0$ for $l\ne l'$.
We can then represent the PHM as $\Xi_{lm}$,
where the indices $l$ and $m$ run over all atomic origins and destinations of particles and holes.
The result is a two-dimensional array, analogous to the maps we obtained in the spatial box partitioning
scheme. \cite{Li2015,Li2016}
To check the equivalence of the two schemes, we compared PHMs calculated using {\tt octopus} and {\tt Gaussian}, and found them to be very similar.

In practice, the value of the PHM can vary significantly between individual pixels. We found it to be advantageous
to introduce a trimming procedure of the data in order to reduce the amplitude of the largest outliers;
this improves the graphical representation without any loss or distortion of physical information.
Defining a cutoff $C = [\max(|\Xi|) + \sigma(\Xi)]/2$,
where $\max(|\Xi|)$ and $\sigma(\Xi)$ are the absolute maximum and the standard deviation of the set of all $\Xi_{lm}$,
the trimmed PHM is
\begin{equation} \label{amplitude}
\textrm{Trim}\left[\Xi_{lm} \right] = \left\{\begin{array}{cr}
\Xi_{lm}, & \left|\Xi_{lm} \right|\leq C  \\
\textrm{sign}(\Xi_{lm})C, & \left|\Xi_{lm} \right| >  C
\end{array} \right. \:.
\end{equation}

As an alternative to a  coarse-grained two-dimensional map, we can visualize the PHM  similar to the way in which excitons
are often represented in the literature:\cite{Rohlfing2000,Ehrhart2014} to plot the
exciton wave function (which also depends on six spatial variables),
one fixes the position of the electron on a selected atomic site and then plots the associated distribution of the hole, or vice versa.

In our case, we plot $\Xi(\bfr_0,\bfr',\Omega_n)$ for a given fixed reference point $\bfr_0$ as a function of $\bfr'$.
The reference point $\bfr_0$ is a certain point of origin of a particle or hole, and the distribution as a function of $\bfr'$ then tells us where
this particle or hole goes. The choice of $\bfr_0$ is in principle arbitrary, but chemical insight will suggest selecting
reference points that sit on certain atomic centers or functional groups of importance within the molecule.
Using appropriate isosurfaces, this then produces direct 3D images of the PHM in a molecule, which, taken together, provide a detailed picture of the
flow of particles and holes between different molecular regions during an excitation.

\subsection{$\rm C_2H_4$--$\rm C_2F_4$ charge transfer complex}

\begin{figure}
\includegraphics[width=\linewidth]{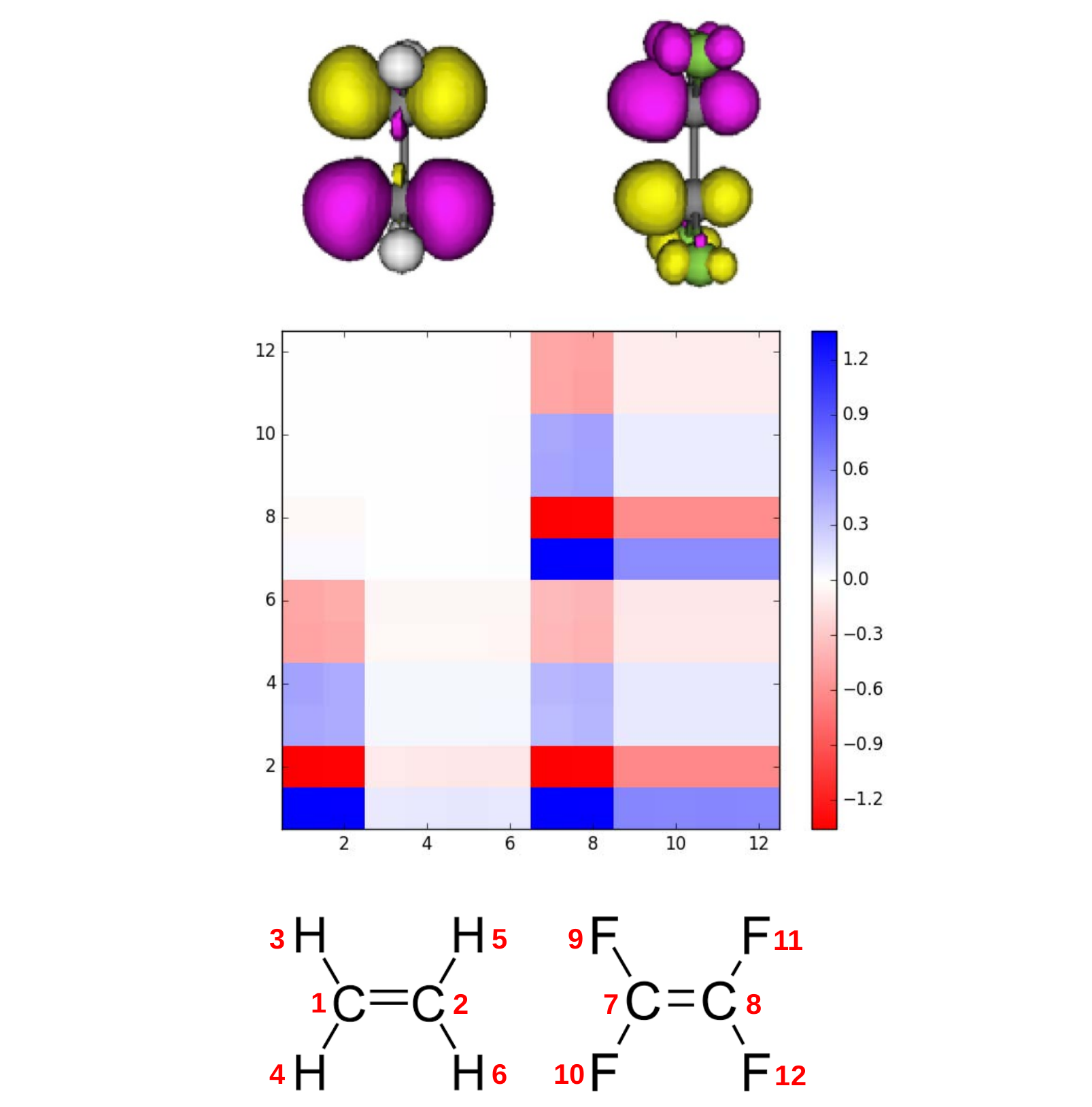}
\caption{Top: transition density of the second excitation of the  $\rm C_2H_4$--$\rm C_2F_4$ dimer at 4 {\AA} separation.
Middle: associated PHM. The axis labels correspond to the numbering of the atomic sites shown in the structure formulas at the bottom.} \label{fig1}
\end{figure}

\begin{figure}
\includegraphics[width=\linewidth]{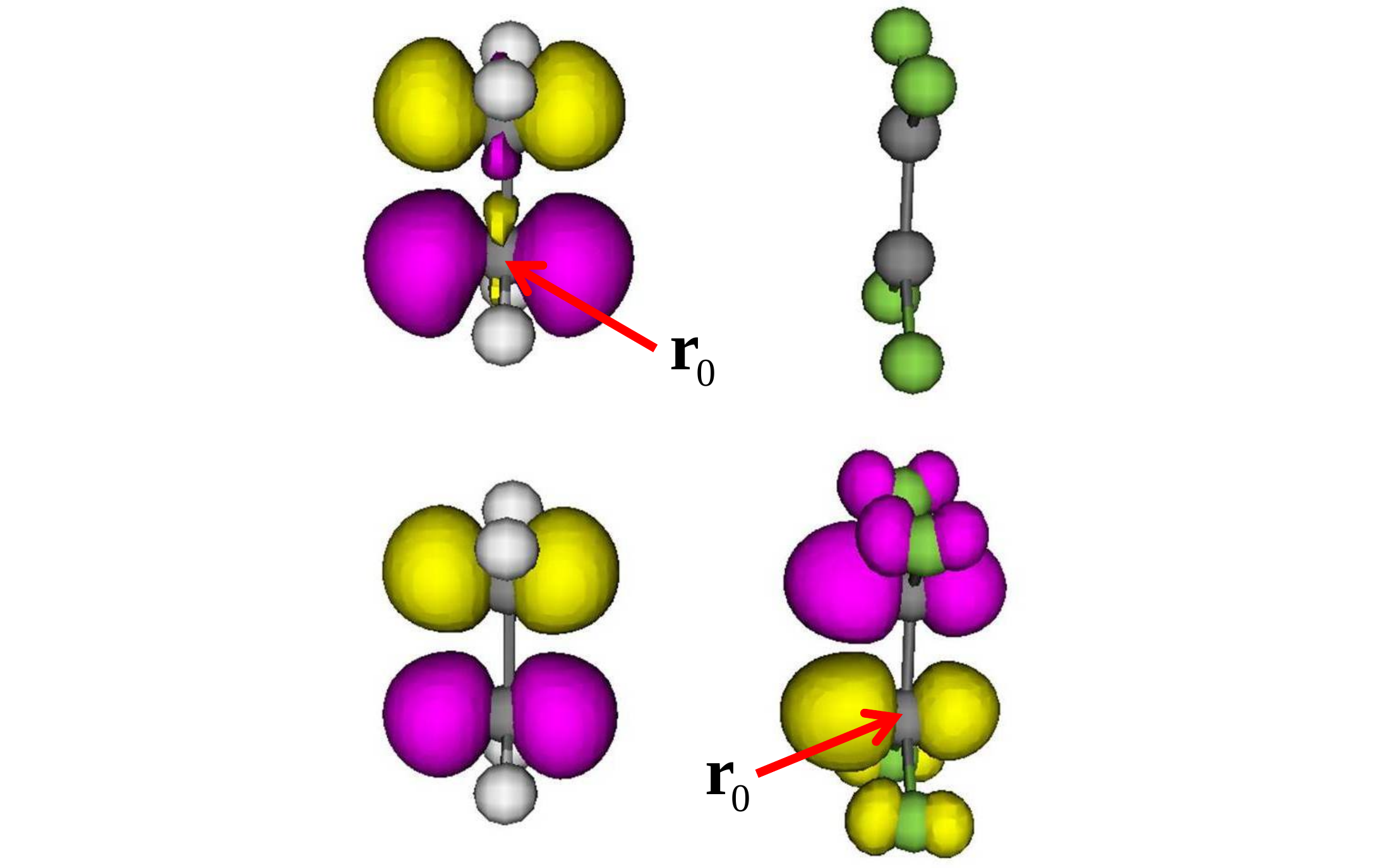}
\caption{Isosurfaces of the PHM $\Xi(\bfr_0,\bfr',\Omega_n)$ for two different reference points $\bfr_0$. The excitation
is the same as in Fig. \ref{fig1}.} \label{fig2}
\end{figure}

We now illustrate these visual representations of the PHM using a simple example.
The ethylene--tetrafluoroethylene dimer has been widely studied as a prototype charge transfer system which provides important
benchmark results for TDDFT. \cite{Dreuw2003,Tawada2004,Neugebauer2006} We here consider the $\rm C_2H_4$--$\rm C_2F_4$ dimer separated
by a distance $R=4$ {\AA}, using the CAM-B3LYP long-range corrected exchange-correlation functional \cite{Yanai2004} and using
the standard 6-31G(d) basis set without geometry optimization. The linear response is calculated in the space of 24
occupied and 96 virtual Kohn-Sham orbitals, using
{\tt Gaussian}. \cite{g09}

Figure \ref{fig1} shows the transition density and the PHM for the second singlet excitation at 7.81 eV; this excitation
has a rather small oscillator strength, which is typical for charge-transfer excitations.
The atomic sites are numbered as follows.\cite{footnote}
1,2: C atoms on $\rm C_2H_4$; 3-6: H atoms on $\rm C_2H_4$; 7,8: C atoms on $\rm C_2F_4$; 9-12: F atoms on $\rm C_2F_4$. This provides the axis labels of the PHM in Fig. \ref{fig1}.

The transition density shows equally pronounced features on both molecules, which
could, in principle, be caused by charge redistributions {\em within} each molecule. However, the PHM makes it very clear
that this is an excitation with a very strong charge-transfer character. The horizontal axis of the PHM indicates
where the electrons (blue) and holes (red) are coming from, and the vertical axis indicates where they are going to.
Thus, the two diagonal blocks, $\mbox{(1-6)}\times\mbox{(1-6)}$ and $\mbox{(7-12)}\times\mbox{(7-12)}$,
indicate charge redistributions within the same molecule, and the off-diagonal blocks,
$\mbox{(1-6)}\times\mbox{(7-12)}$ and $\mbox{(7-12)}\times\mbox{(1-6)}$,
indicate charge transfer from the left to the right fragment and vice versa.

As the PHM in Fig. \ref{fig1} shows, the upper left off-diagonal block is empty, which means that there is no transfer
from $\rm C_2H_4$ to $\rm C_2F_4$. The lower right off-diagonal block, on the other hand, gives evidence of a very strong
charge transfer from $\rm C_2F_4$ to $\rm C_2H_4$. The varying degrees of lightness of the individual pixels give evidence
for the contributions of the atomic orbitals associated with the C, H, and F atoms to the charge transfer between the molecules.
The C-C transfer is most pronounced, whereas the H-F transfer appears to be weakest.

Figure \ref{fig2} shows the isosurfaces of the PHM $\Xi(\bfr_0,\bfr',\Omega_n)$ for two reference points $\bfr_0$.
The top part shows the PHM for $\bfr_0$ on a C atom of $\rm C_2H_4$: the PHM only extends within the $\rm C_2H_4$ molecule itself.
Since $\bfr_0$ is a specific origin of electrons and holes, and $\bfr$ is their destination, this shows clearly that there
is no charge transfer from the left to the right molecule. On the other hand, choosing $\bfr_0$ to lie on a C atom of
$\rm C_2F_4$ produces a PHM which extends over {\em both} molecules, which indicates a strong charge transfer from the right
to the left molecule.

\section{Conclusion}

In this paper we have presented a rigorous derivation of the particle-hole map, starting from a
two-body reduced transition density matrix (in the particle-hole channel) and projecting onto a set of occupied canonical
single-particle orbitals. Apart from justifying our earlier {\em ad hoc} definition, this new derivation
shows that the PHM is an object that is very close to a physical observable. The additional projection
step is straightforward to do in practice, and lends itself to generalization
by using many-body wave functions and Dyson orbitals.

We then showed that the PHM can be quite easily implemented using the output of TDDFT calculations from standard
quantum chemistry codes.
The PHM can be represented as a two-dimensional map whose labels are obtained by
consecutively numbering the individual atoms, or one can plot isosurfaces for a given reference point $\bfr_0$.
We showed that both representations yield detailed insight into the inner mechanisms of excitation processes
that are difficult to obtain in other ways.

In addition to the molecular charge-transfer excitations discussed here, the PHM can be used
in extended systems and periodic solids. It can, in principle, also be generalized to the spin-dependent
case, which would provide novel insight into triplet excitations or magnetization dynamics. Work along
these lines is in progress.

\acknowledgments

This work was supported by National Science Foundation Grant No. DMR-1408904.
We thank Evert Jan Baerends for valuable discussions.


\begin{thebibliography}{99}
\bibitem{footnote}
We chose a numbering scheme that groups similar atoms together over one dictated by
the spatial connectivity within the molecules. We found this to be preferable for the small system considered here.
In general, the atomic labeling should be such that the resulting maps are as easy to read and interpret as possible;
this may require some compromise in practice.
\end{thebibliography}
\end{document}